\documentclass[sigconf,authorversion]{acmart}

\usepackage{booktabs} 
\usepackage[utf8]{inputenc}
\usepackage{xcolor} 	

\usepackage{pgfplots}
\pgfplotsset{width=\columnwidth, compat=1.8}
\usepgfplotslibrary{statistics}

\begin{document}

\copyrightyear{2018}
\acmYear{2018}
\setcopyright{acmlicensed}     
\acmConference[SIGCSE '18]{The 49th ACM Technical Symposium on Computer Science Education}{Feb. 21--24, 2018}{Baltimore, MD, USA}
\acmPrice{15.00}
\acmDOI{10.1145/3159450.3159561}
\acmISBN{978-1-4503-5103-4/18/02}

\title{Timely Feedback in Unstructured Cybersecurity Exercises}



\author{Jan Vykopal}
\orcid{0000-0002-3425-0951}
\affiliation{%
  \institution{Masaryk University, Institute of Computer Science}
  \streetaddress{Botanická 68a}
  \city{Brno} 
  \state{Czech Republic}
  \postcode{60200}
}
\email{vykopal@ics.muni.cz}

\author{Radek Ošlejšek}
\orcid{0000-0002-0562-6892}
\affiliation{%
  \institution{Masaryk University, Faculty of Informatics}
  \streetaddress{Botanická 68a}
  \city{Brno} 
  \state{Czech Republic} 
  \postcode{60200}
}
\email{oslejsek@fi.muni.cz}

\author{Karolína Burská}
\affiliation{
  \institution{Masaryk University, Faculty of Informatics}
  \streetaddress{Botanická 68a}
  \city{Brno} 
  \state{Czech Republic} 
  \postcode{60200}
}
\email{burska@mail.muni.cz}

\author{Kristína Zákopčanová}

\affiliation{%
  \institution{Masaryk University, Faculty of Informatics}
  \streetaddress{Botanick8 68a}
  \city{Brno} 
  \country{Czech Republic}}
\email{zakopcanova@mail.muni.cz}

\renewcommand{\shortauthors}{J. Vykopal et al.}

%
%

\begin{abstract}

Cyber defence exercises are intensive, hands-on learning events for teams of professionals who gain or develop their skills to successfully prevent and respond to cyber attacks. The exercises mimic the real-life, routine operation of an organization which is being attacked by an unknown offender. Teams of learners receive very limited immediate feedback from the instructors during the exercise; they can usually see only a scoreboard showing the aggregated gain or loss of points for particular tasks. An in-depth analysis of learners' actions requires considerable human effort, which results in days or weeks of delay. 
The intensive experience is thus not followed by proper feedback facilitating actual learning, and this diminishes the effect of the exercise.

In this initial work, we investigate how to provide valuable feedback to learners right after the exercise without any unnecessary delay. 
Based on the scoring system of a cyber defence exercise, we have developed a new feedback tool that presents an interactive, personalized timeline of exercise events. 
We deployed this tool during an international exercise, where we monitored participants' interactions and gathered their reflections. The results show that learners did use the new tool and rated it positively. 
Since this new feature is not bound to a particular defence exercise, it can be applied to all exercises that employ scoring based on the evaluation of individual exercise objectives. As a result, it enables the learner to immediately reflect on the experience gained.
\end{abstract}

\maketitle


\section{Introduction}

Cyber attacks threatening ICT infrastructure have become routine. 
Their intensity and complexity are growing with the increasing number of interconnected devices exposed to attackers and the influx of new vulnerabilities being revealed each year.
Unfortunately, there is a significant global shortage of cybersecurity workers equipped with the skills necessary for preventing or responding to the attacks~\cite{GISWS-2015}. 

Cyber defence exercises (CDX)~\cite{exercise-taxonomy} represent a popular type of training that aims to fill this skill gap. They enable participants to experience cyber attacks first-hand with real-life limitations, including a lack of information and resources, the need for communication and making decisions under stress. 

CDX are usually intensive, short-term events lasting several days. Tens to hundreds of professional learners participate and are grouped in teams. The target groups are administrators of ICT systems, incident responders, and security managers. 
The exercises deliver rich immediate experience of ongoing attacks and the opportunity to practice crisis procedures and techniques.

In contrast to structured, step-by-step hands-on training guided by an instructor, teams of learners have to figure out all the issues on their own, in the order they agree on, within the team. These settings simulate real operation but prevent any direct feedback from instructors (exercise organizers). Learners can only presume what they did was correct, what worked and what did not. The only feedback they are given during the exercise is often through an exercise score with no further details about score breakdown. Some exercise organizers provide technical reports after the exercise, which reveal some details highlighting important moments from the perspective of a particular team of learners. The after-action report is sometimes also complemented by a short workshop, which provides an opportunity to discuss the content of the exercise with the instructors in person. Nevertheless, all these methods of feedback are delivered with a significant delay after the actual exercise because they require preparation from the instructors that cannot begin before the end of the exercise.

In this paper, we study whether learners benefit from simple, but individualized feedback provided just after the end of a two-day intensive exercise. In our study, each team was provided with an interactive timeline of its score development during the exercise, with important events emphasized. The timeline was generated automatically from data stored by an existing scoring system. All interactions of exercise participants (mouse clicks and movements) were logged with the scoring timeline. After that, participants were asked to fill out short evaluation questionnaire. 
The data and answers we obtained show that learners valued the feedback, even though they still lack more details about particular events.

\section{State of the art}

Research on providing feedback in complex and unstructured cybersecurity exercises is very scarce. The following overview is therefore based not only on a review of academic literature but also on technical reports published by organizers and the experience of authors who participated in several CDXs.


One of the world's largest exercise is Locked Shields~\cite{nato-exercises}, which is organized annually by the NATO Cooperative Cyber Defence Centre of Excellence in Tallinn, Estonia. 
Immediate brief feedback to learners is provided at a so-called "hot wash-up" session right after the exercise since any time lag will diminish the learning impact~\cite{maennel-thesis}. Educators provide a summary of the exercise's progression and comment on key moments that drove it. More comprehensive and detailed feedback is available only at a workshop which is held a month later, and furthermore, not all learners come to this event.

Another international exercise organized by NATO is Cyber Coalition. Very brief feedback is provided a day later in-person at a hot wash-up session. More specific information is available only in the form of an after-action report at a workshop, which takes place a month later~\citep{cyber-coalition}. A very similar type of feedback, with an even longer delay is provided in Cyber Europe, another international exercise organized by ENISA.~\citep{enisa}


Gran{\aa}sen and Andersson~\cite{Granasen2016} focused on measuring team effectiveness in CDXs. They thoroughly analysed system logs, observer reports, and surveys collected during Baltic cyber shield 2010, a multi-national civil-military CDX. 
They concluded that these multiple data sources provided valuable insight into the exercise's run. However, they did not mention how to use these data for providing feedback. The only feedback provided to learners was during a virtual after-action review the day after the exercise, where only the leaders of learners' and organizers' teams summarized and discussed their experience.

Henshel et al.~\cite{7795423} also focused on the assessment model and metrics for team proficiency in CDXs. They analysed learners' and observers' input from surveys and intrusion detection logs from the Cyber Shield 2015 exercise. Similarly to the Baltic cyber shield exercise, the feedback provided to learners was not based on an analysis of acquired data since the analysis was done manually and required significant human effort.


Since existing CDXs do not provide any timely and personalized feedback to the learners, we also mention a feedback-related study in learning, which does not concern CDXs.   
Gibbs and Taylor~\cite{gibbs-feedback} focused on the theory that personalized feedback does not hold such importance or a value compared to its time-consumption for the instructor. 

\section{Experiment setup}

In this experiment, we studied the behaviour and interactions of participants at a complex cyber defence exercise held on May 23--24, 2017 at Masaryk University, Brno, Czech Republic. 
The exercise is focused on defending critical information infrastructure (particularly railway infrastructure administration) against skilled and coordinated attackers. 

First, the learners got access to the exercise infrastructure to become familiar with virtual hosts in their network. Then, they took part in an intensive exercise where they faced challenges posed by simulated attackers and legitimate users. Right after the end of the exercise, the learners were asked to express their immediate impressions about the exercise in a short post-exercise survey. After a short break, a timeline depicting the score of each team was presented in the exercise portal. Finally, the learners were asked to evaluate the timeline via a very short questionnaire. Time allocated for each phase of the experiment is shown in Table~\ref{table:experiment-design}.

\begin{table}
\begin{center}
\caption{Phases of the exercise with time allocation}
\begin{tabular}{|c|l|r|c|}
\hline
Order & Phase & Duration & Day \\ \hline \hline
1 & Exercise familirization & 3 hrs & 1 \\ \hline
2 & Actual exercise & 6 hrs & 2 \\ \hline
3 & Post-exercise survey & 5 mins & 2 \\ \hline
{\color{gray} 4} & {\color{gray} Break} & {\color{gray} 25 mins} & {\color{gray} 2}\\ \hline
5 & Scoring timeline interaction & 10 mins & 2\\ \hline
6 & Scoring timeline survey & 5 mins & 2\\ \hline
{\color{gray} 7} & {\color{gray} Quick exercise debriefing} & {\color{gray} 15 mins} & {\color{gray} 2}\\ \hline 
\end{tabular}
\label{table:experiment-design}
\end{center}
\end{table}

\subsection{Exercise participants}

\begin{figure}[ht]
  \centering
  \includegraphics[width=\columnwidth,draft=false]{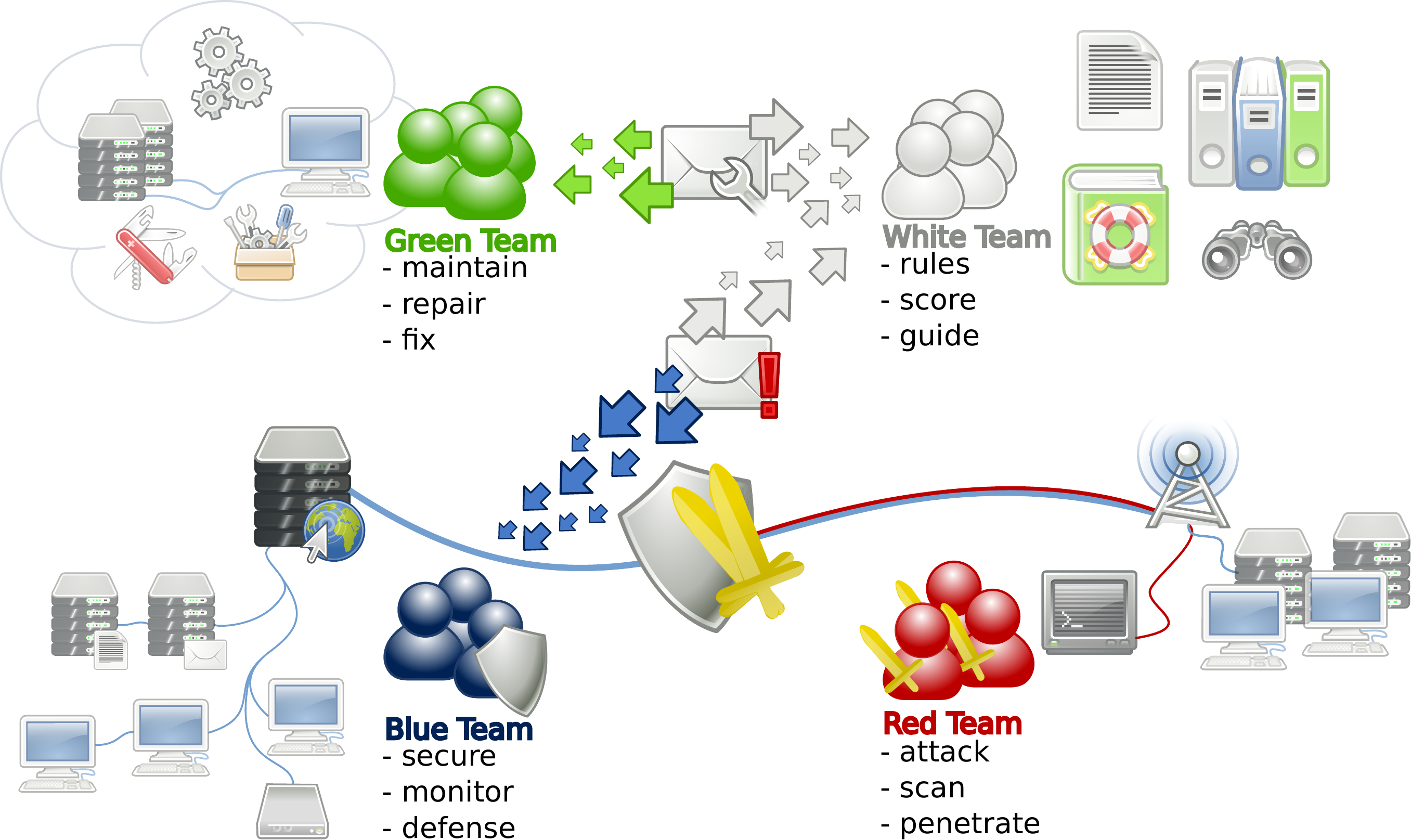} 
  \caption{Exercise participants, their interactions and tasks.}	\label{fig:team_roles}
\end{figure}

The experiment involved a Red vs. Blue exercise with 40 participants working in an emulated ICT infrastructure. The structure of the exercise is inspired by the Locked Shield exercise~\cite{nato-exercises,cdx-kypo-fie2017}. The participants were divided into four groups according to their role and tasks in the exercise. Their interactions are depicted in Figure~\ref{fig:team_roles}.

Twenty professional learners formed five \emph{Blue teams} (T1--T5) which were put into the role of emergency security teams sent into five organizations to recover compromised networks. Each team of 4 learners was responsible for securing the compromised networks, dealing with the 
attacks, collaborating with other emergency teams, and collaborating with the coordinator of the operation and media representatives. 
They had to follow the exercise's rules and local cybersecurity law. 
Each team represented one real cybersecurity response team from one country in Central Europe.

All attacks against infrastructure defended by Blue teams were conducted by \emph{Red team}. This team consisted of cyber security professionals 
who carefully followed a predefined attack scenario to equally load the Blue teams. This means they should not use any other arbitrary means of attack against the Blue teams. 
Based on the success of attacks, the Red team assigns penalty points to the Blue teams
since the amount of points is based on non-trivial factors that need expert review.

Exercise managers, referees, organizers, and instructors worked in \emph{White team}. 
During the exercise, this team assigns tasks (called injects) to the Blue teams and thus simulates the requests of many entities, such as legitimate users of the defended organization, the operation coordinator which needs situational reports, media inquiries, and law enforcement agencies. Then the White team assesses the promptness and quality of a Blue teams's reactions to these tasks and assigns penalties and points.

Finally, the \emph{Green team} is a group of operators and system administrators responsible for the exercise infrastructure. They have full access to the exercise network so they can provide assistance to Blue teams in trouble in an exchange for penalty points.  

\subsection{Exercise phases}

Before the actual exercise (Phase 1 in Table~\ref{table:experiment-design}), learners are provided with a background story to introduce them to the situation before they enter the compromised networks. Then they access their part of the emulated network for 3 hours to get familiar with the exercise infrastructure.
This is very important since the exercise is not set in a known environment and learners have no previous knowledge about who is who in the fictitious scenario (e.\,g., users in their organization, a popular news portal, superordinate security team). 

The exercise (Phase 2) is driven by a detailed scenario which includes the actions of attackers (Red team) and assignments for the defenders prepared by the organizers (White team). Attackers exploit specific vulnerabilities left in the compromised network in a fixed order. This follows a common attack life cycle in a critical information infrastructure. 
On top of that, learners should also answer media inquires and requests from users doing their routine job in the defended network. The performance of each Blue team is scored based on successful attacks or their mitigation, the availability of specified critical services and the quality of reporting and communication. The score is either computed automatically from events, processed by the logging infrastructure (e.\,g., a penalty for inaccessible services) or entered manually (e.\,g., attacks completed by the Red team). An aggregated score is shown to participants in real-time. Table~\ref{tab:scoring-table} shows the structure of the scoreboard and the values of aggregated score.

\begin{table}[ht]
\caption{The scoreboard presented to the learners during the exercise.}
\label{tab:scoring-table}
\begin{minipage}{\columnwidth}
\begin{center}
\begin{tabular}{l|rrrrr|c}
\toprule
Team & Services & Attacks & Injects & Users & Access & Total\\
\hline \hline
{\bf T1} & 91,843 &  -8,500 &  9,000 & -1,100 &      0 & 91,243 \\
{\bf T5} & 92,230 &  -5,000 &  3,600 &   -400 &      0 & 90,430 \\
{\bf T2} & 81,280 & -10,750 &  6,425 & -4,000 &      0 & 72,955 \\
{\bf T4} & 74,518 & -11,000 &  6,650 &      0 & -4,000 & 66,168 \\
{\bf T3} & 85,756 & -12,000 &  2,475 & -1,700 & -9,500 & 65,031 \\
\bottomrule
\end{tabular}
\bigskip\\
\footnotesize Note: Teams are sorted according to their final score. \\ ''Injects'' is an abbreviation for communication injects of the White team.
\end{center}
\end{minipage}
\end{table}

Immediately after the end of the exercise, we asked the learners to evaluate the exercise and their experience by rating several statements using the Likert scale (Phase 3).

\subsection{Scoring timeline application} 

\begin{figure*}[ht]
  \centering
  \includegraphics[width=\linewidth]{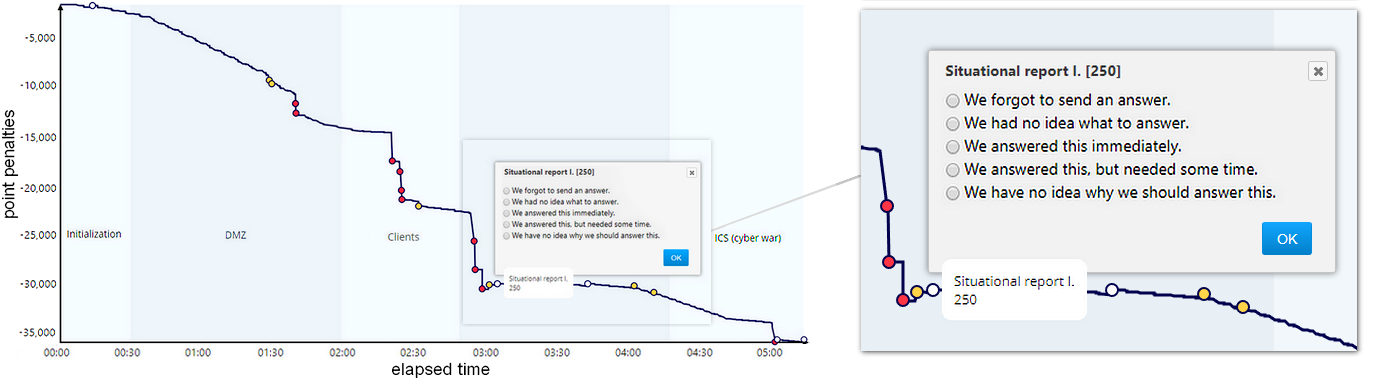} 
  \caption{A screenshot of a scoring timeline providing personalized feedback for each team right after the exercise.}			     \label{fig:timeline}
\end{figure*}

After a break (Phase 4), the score acquired by each team during the exercise was presented to the Blue teams in the form of an interactive application, as shown in Figure~\ref{fig:timeline} (Phase 5). It provided automatically generated personalized feedback. Members of each team could only see their own scoring timeline with individual exercise events. 

The initial score of each team is 100 000 points. In the graph, the main, predominantly descending line represents the development of a team's total score over time. 
The score is computed from penalties and awarded points that were either recorded automatically for the inaccessibility of required network services or assigned manually by the Red or White team. The colourful dots are interactive and they are related just to the manual rating. The red dots represent Red team penalties, white dots represent the rating of communication injects by the White team, yellow dots indicate the rating of user simulated injects by the White team, and grey dots indicate requests for assistance from the Green team (to grant temporary remote access to a machine or revert to the initial state). Each dot contains textual information that specifies the reason for the rating. This information is shown in each dot's tooltip.



Learners were able to provide us with their reflection on their penalty and awarded points very easily by clicking on the coloured dots and choosing one of predefined options (Phase 5; see right-hand side of Figure 2), e.\,g., whether they recognized the attack or not, or why they did not respond to the inject of the White team. Moreover, all scoring timeline interactions, including mouse clicks, mouse movements, and selected options were logged. This data, together with answers from a short survey on the scoring timeline (Phase 6) were used to evaluate the usefulness of the timely feedback. 

Finally, representatives of the Red and White team provided a short debrief of the exercise (''hot wash-up'', Phase 7) from their perspective. 
They highlighted important breaking points of exercise and pointed out exemplary or interesting decisions and actions took by Blue teams. 
This part is mentioned here only for completeness, no input from learners was required for the experiment.

\section{Results}

\subsection{Post-exercise survey} \label{sec:results-quality}

\begin{table*}[ht]
\caption{Team statistics for the post-exercise and scoring timeline surveys}
\label{tab:questionnaire}
\begin{minipage}{\textwidth}
\begin{center}
\begin{tabular}{lp{7cm}llllll}
\toprule
& {\bf Statement} & {\bf Team 2} & {\bf Team 1} & {\bf Team 5} & {\bf Team 4} & {\bf Team 3} & {\bf Average}\\
\hline \hline
E1: & My knowledge and skills were sufficient.    & 3.75 & 3    & 2.6 & 2.5  & 2 & 3.05\\
E2: & I found exercise difficult for me.          & 3.33 & 3    & 4   & 4.25 & 5 & 3.8\\
E3: & Exercise was well organized and structured. & 2.66 & 3.25 & 3.3 & 4    & 5 & 3.75\\
E4: & Exercise was beneficial and useful to me.   & 2.66 & 3.5  & 4   & 4.5  & 5 & 3.85\\
\hline
F1: & The scoring timeline of my team displayed after the end of the exercise provided useful feedback. & 3.25 & 2.5 & - & 4.25 & 3.66 & 3.53\\
F2: &  Do you have any comments on the scoring timeline? & D & M & - & M & M & \\
\bottomrule
\end{tabular}
\bigskip\\
\footnotesize1 = strongly disagree, 5 = completely agree, D = there was a delay in inserting points by a Red and White team, M = add more details about the depicted events
\end{center}
\end{minipage}
\end{table*}

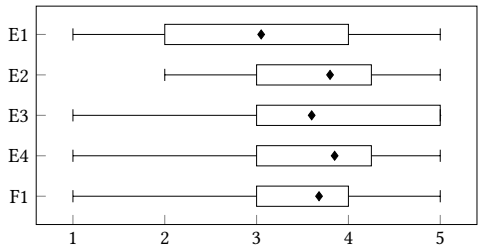
\begin{figure}[ht]
\centering
  \begin{tikzpicture}[scale=0.85]
  \begin{axis}
    [
      boxplot/draw direction=x,
      ytick={1,2,3,4,5},
      yticklabels={F1, E4, E3, E2, E1},
      xtick pos=left,
      ytick pos=left,
      height=5cm,
    ]
    \addplot+[
    	color=black,
    	boxplot prepared={
        	box extend=0.5, whisker extend=0.3,
            upper quartile=4, lower quartile=3, upper whisker=5, lower whisker=1, average = 3.68},
    ] coordinates {};
    \addplot+[
    	color=black,
    	boxplot prepared={
        	box extend=0.5, whisker extend=0.3,
            upper quartile=4.25, lower quartile=3, upper whisker=5, lower whisker=1, average = 3.85},
    ] coordinates {};
    \addplot+[
    	color=black,
    	boxplot prepared={
        	box extend=0.5, whisker extend=0.3,
            upper quartile=5, lower quartile=3, upper whisker=5, lower whisker=1, average = 3.6},
    ] coordinates {};
    \addplot+[
    	color=black,
    	boxplot prepared={
        	box extend=0.5, whisker extend=0.3,
            upper quartile=4.25, lower quartile=3, upper whisker=5, lower whisker=2, average = 3.8},
    ] coordinates {};
    \addplot+[
    	color=black,
    	boxplot prepared={
        	box extend=0.5, whisker extend=0.3,
            upper quartile=4, lower quartile=2, upper whisker=5, lower whisker=1, average = 3.05},
    ] coordinates {};
  \end{axis}
\end{tikzpicture}
\caption{Distribution of all answers to E1 -- E4 and F1.}
\label{fig:boxplot-questionnaire}
\end{figure}

The post-exercise survey was focused on general qualitative aspects of the exercise. Relevant statements to this study are listed in Table~\ref{tab:questionnaire} and marked as E1--E4 for further reference. The statistical distribution of individual answers across all teams is depicted in Figure~\ref{fig:boxplot-questionnaire}.

We collected answers from all 20 participants. However, four of them did not provide their identification and so their answers were omitted from the team statistics. Team values were computed as the average from a five-point Likert scale answers (1~=~strongly disagree, 5~=~completely agree). The \emph{Average} column represents the average value across all answers regardless of teams. 

The aim of E1 and E2 was to reveal the level of expertise of individual teams and the difficulty of the exercise. Individual answers to E1 significantly varied, which indicates that learners had significantly different expertise. However, the average values calculated for each team reveal that the exercise was well balanced with no extremely weak or strong team. The answers to E2 indicate that the overall difficulty of the exercise was considered as rather high.

Two statements in the questionnaire, E3 and E4, were focused on satisfaction with the organizational aspects and usefulness of the exercise. Both statements brought very similar answers. Teams that were more satisfied with the organization also considered the exercise more beneficial. Even though the opinion differed across the teams, learners considered the exercise rather beneficial and well organized in general. 

\subsection{Scoring timeline interaction}

\begin{figure}[ht]
  \centering
  \includegraphics[width=\columnwidth]{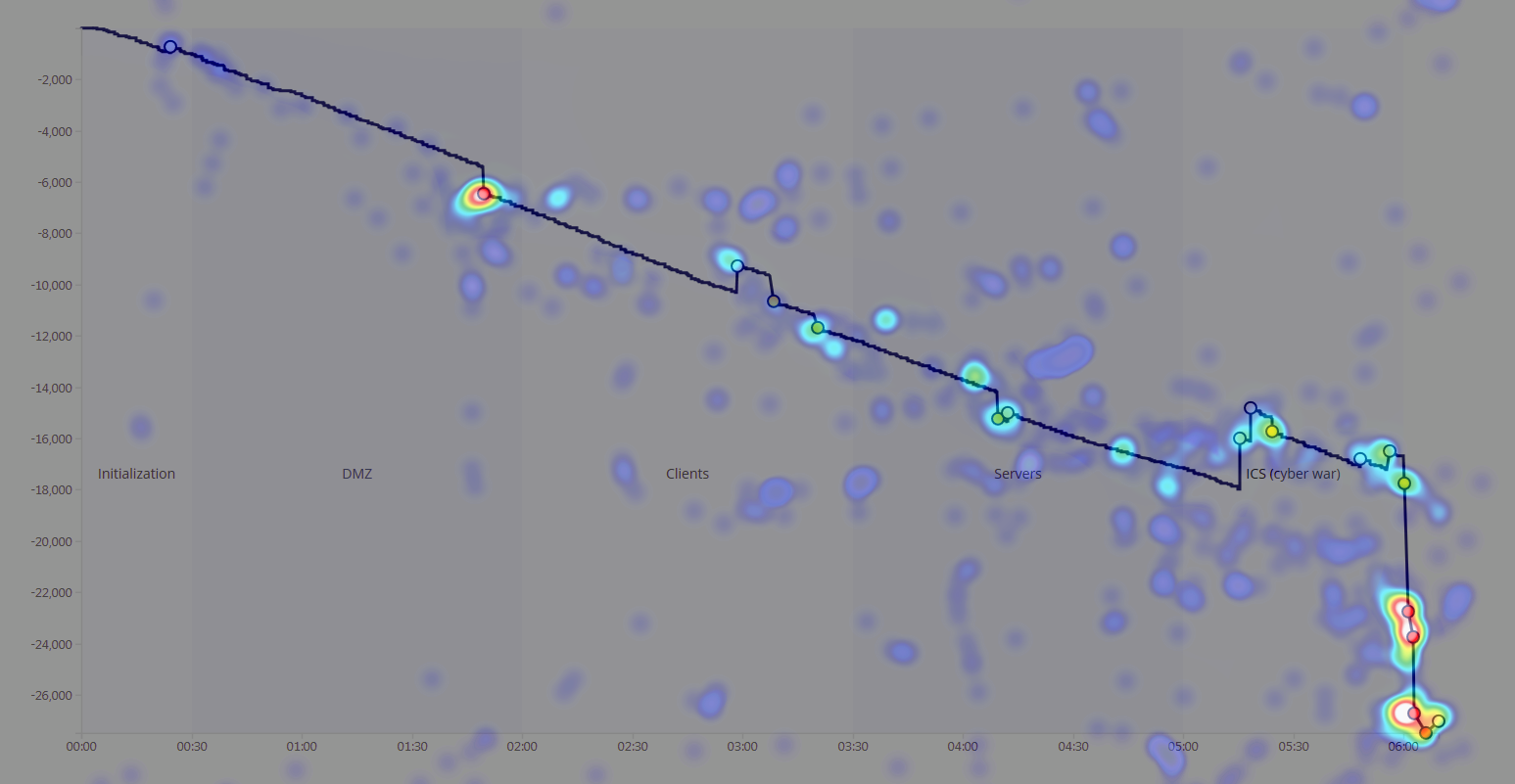}
  \caption{An example of a heatmap of mouse movements and clicks on a screen of a scoring timeline.} 

  \label{fig:timeline-heatmap}
\end{figure}

In order to evaluate the scoring timeline, we were actively recording learners' interactions with a tool.
We obtained data from 18 learners, 2 learners were missing due to technical issues. The data consisted of 2,994 individual low-level events (mouse clicks, mouse hovers, etc.). Moreover, we recorded heatmaps of mouse positions on the screen (see an example in Figure~\ref{fig:timeline-heatmap}). Putting this data together, we were able to estimate the focus of the learners during their exploration of the scoring timeline. A deeper analysis did not reveal any preference patterns in the sense that learners would be more interested in some kind of objectives, such as Red team attacks, White team injects, or penalties from later critical phases of the exercise. On the contrary, it seems that the learners were interested in all the penalties and awards roughly the same.

Analyses of timestamps revealed that the time spent by individual learners with the scoring timeline ranged between 1m~22s and 8m~27s
It is worth noting that there was no significant difference between teams. They spent approximately 3 to 5 minutes with the feedback application on average. We also did not find any relation between the time spent with the timeline and the willingness to provide their reflection on particular penalties or awards. Many learners just spent a long time with only a passive exploration of the timeline. 

Seven learners from four teams also gave us an active reflection to penalties in addition to passively exploring the scoring timeline. The values in Table~\ref{tab:feedback} represent the numbers of collected answers per team and objective type. Blue team 1 is omitted from the table because we got no data from them. These results are discussed in Section~\ref{sec:discussion}.

\begin{table}[ht]
\caption{Numbers of responses of each team to 
objectives.}
\label{tab:feedback}
\begin{minipage}{\columnwidth}
\begin{center}
\begin{tabular}{lccccc}
\toprule
& \multicolumn{4}{c}{\bf Teams} & \\
{\bf Objectives} & T2 & T3 & T4 & T5 & $\sum$ \\
\hline \hline
Red team attacks & 7 & 13 & 5 & 1 & 26\\
Users injects & 5 & 7 & 0 & 1 & 13\\
Communication injects & 0 & 5 & 0 & 2 & 7\\
Green team assistance & 0 & 4 & 2 & 0 & 6\\
\bottomrule
\end{tabular}
\end{center}
\end{minipage}
\end{table}

\subsection{Scoring timeline survey} \label{sec:results-timeline} 

The usefulness of the feedback provided via the scoring timeline was evaluated with a short survey. We got answers from 13 learners (out of 20) because Blue team 5 did not respond at all. The statements and their team statistics are shown in Table~\ref{tab:questionnaire} under the labels F1 and F2. 

\begin{figure}[ht]
  \centering
  \begin{tikzpicture}[scale=0.85]
  \begin{axis}
    [
      boxplot/draw direction=x,
      ytick={1,2,3,4},      
      yticklabels={Team 3, Team 4, Team 2, Team 1},
      xtick pos=left,
      ytick pos=left,
      height=4cm,
    ]
    \addplot+[
    	color=black,
    	boxplot prepared={
        	box extend=0.5, whisker extend=0.3,
            upper quartile=4, lower quartile=3.5, upper whisker=4, lower whisker=3, average = 3.66},
    ] coordinates {};
    \addplot+[
    	color=black,
    	boxplot prepared={
        	box extend=0.5, whisker extend=0.3,
            upper quartile=5, lower quartile=3.75, upper whisker=5, lower whisker=3, average = 4.25},
    ] coordinates {};
    \addplot+[
    	color=black,
    	boxplot prepared={
        	box extend=0.5, whisker extend=0.3,
            upper quartile=4, lower quartile=3, upper whisker=4, lower whisker=3, average = 3.4},
    ] coordinates {};    
    \addplot+[
    	color=black,
	    boxplot prepared={
        	box extend=0.5, whisker extend=0.3,
            upper quartile=3.25, lower quartile=1.75, upper whisker=4, lower whisker=1, average = 2.5},
    ] coordinates {};
  \end{axis}
\end{tikzpicture}
  \caption{Distribution of answers to F1. Teams are sorted according to the final score (the best on the top).}
  \label{fig:f1}
\end{figure}
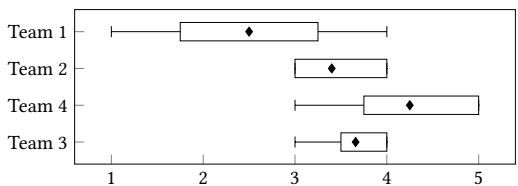

The data shows that the usefulness of the feedback can be considered to be "rather useful" (the average value across all the teams is 3.53). A detailed distribution of answers depicted in Figure~\ref{fig:f1} reveals that the most successful team considered the feedback less useful than other teams.

Answers to the open question F2 provided four comments on possible improvements. Three teams requested more details about penalties. They would appreciate knowing "how it happened" in addition to "what happened" (marked as \emph{M} in Table~\ref{tab:questionnaire}). One comment objected that "some attacks were happening way sooner in reality than on the timeline" (marked as \emph{D} in Table~\ref{tab:questionnaire}). This is true because attack penalties were inserted by the Red team manually with some delay.

\section{Discussion} \label{sec:discussion}

%

The validity of the timely feedback evaluation would be affected by dissatisfaction with the exercise so the learners would not be interested in the feedback at all or they would provide distorted data. Dissatisfaction can be caused by poor organization, unfulfilled expectations, or by a disparity between the difficulty of the exercise and the knowledge of learners. In general, this exercise was considered slightly difficult and was also assessed positively (see Section~\ref{sec:results-quality}). Although the exercise was attended by skilled teams whose members had previous experience with similar exercises, they still considered it rather difficult, challenging and useful. Therefore, we believe that conclusions drawn from the analysis based on this particular exercise and its participants are plausible.
In the following part, we put together the results of the post-exercise survey, the scoring timeline survey, and the exploration of the  timeline and analyse their mutual relationships. 



\paragraph{Teams sought out feedback} 

A deeper analysis of learners' interactions with the scoring timeline shows that all teams were using it intensively, regardless of what reflection they provided in the scoring timeline survey. All teams explored all the penalties depicted in their timeline. They also gave us an evaluation of the majority of displayed objectives. The only exception was Team 4, probably due to technical issues. Team 5, which did not respond to the scoring timeline survey at all, still explored their timeline actively. Team 1 rated the scoring timeline as less useful than other teams. It was the most successful team according to the final score and, therefore, the feedback may not have been so interesting for them because they might have already known about their failures. However, even this team was interested in the timely feedback because they explored the scoring timeline for the longest.

\paragraph{A need for more detail}

Answers to the open question F2 confirmed our assumption that more precise and more detailed information provided by the timely feedback makes the feedback even more attractive to learners. Nevertheless, the need for more detail can also be indirectly inferred. For example, a further analysis of the White team's injects (users and communication injects) indicates that teams often underestimated time-dependant response to this type of "soft" request. Teams could either consider these requests "annoying" and not so important in comparison with attacks or they could just be too busy with the attacks, for instance. A deeper understanding of their behaviour also requires collecting more detailed and better structured data from the timely feedback.

\paragraph{Benefits for instructors} 

Although the scoring feedback was intended primarily for learners, the previous discussion shows that it is very valuable also for organizers and educators who can learn about the exercise and then fine tune its parameters, e.\,g., by better scheduling of the White team's injects with respect to the attacks. Since learners are not usually aware of this value, they do not make more effort than necessary into providing quality and valuable information for instructors. However, if providing reflections is intuitive and quick for them, they are motivated to use it. Furthermore, since the feedback is generated automatically, organizers can get valuable data without any additional effort.

\paragraph{Limitations of the study}

This study is limited in two respects. First, data were obtained from a single exercise with a relatively small group of participants. The reason is that the organization of such a complex exercise is expensive and time-consuming and it is organized rarely for only a narrow group of experts. Nevertheless, we consider the sample to be representative because the learners were highly qualified experts, often with experience in similar exercises and the exercise received a positive rating from them.
Second, the timeline evaluation questionnaire was simple and freely structured. However, this was intentional since the survey took place at the end of an exhausting two-day exercise and thus a more sophisticated questionnaire might not have guaranteed more precise results.

\section{Conclusions and Future work}
To best of our knowledge, this paper is the first attempt to study the means of providing feedback to learners participating in cyber defence exercises. The literature review and our own experience 
showed that feedback provided in state-of-the-art exercises is very limited or delayed.
The most often used method is an exercise scoreboard displayed to the learners throughout the exercise. 
Another method is a short verbal evaluation by exercise observers or organizers right after the end the exercise or its phases. The last commonly used method is after-action reports highlighting key conclusions from a laborious manual analysis of heterogeneous data acquired during the exercise (survey, written communication, scoring and monitoring logs, and checks).

The lack of timely feedback results in the learners having limited opportunity to learn from their experience. They undergo numerous real-life situations during the exercise, but they are not supplied with an explanation why such situations occur. We therefore complemented these means by a novel approach which provides feedback to learners right after the end of the exercise with no additional effort required from educators. Learners can explore a scoring timeline depicting increases and decreases in their team's score and display details about individual events (name and type of the exercise objective, and the number of awarded points or penalty points). 
Also, they have an opportunity to provide their reflections to educators and indicate their awareness about a particular objective and its solution. The exploration and interaction with the scoring timeline enables learners to reflect on their experience and thus strengthen the learning impact of the exercise.

In order to evaluate this approach, we ran an experiment involving a two-day, complex cyber defence exercise with 24 objectives, and 20 professional learners from five security teams. The results, based on an analysis of user surveys and interactions with the new tool, suggest that learners welcomed the new feature even though the feedback was mined automatically and thus provided a very limited level of detail about particular events. 

The experiment also outlined directions for future work. First, learners would appreciate more detail about a particular event in the timeline. 
This can be easily done by adding a detailed description of the objective related to the event. 
Another option is to extend the scoring application so that instructors can not only assign points, but also provide a comment on each exercise objective.
Second, the timeline could be enriched with a display of the relationship between all exercise objectives. This would highlight that an event that the team may not have been aware of was caused by several previous events they encountered. The context can be then built not only from a time perspective, but also from the topology of the exercise network. Providing information on which host or service was affected by the particular event may help learners to recall the particular situation and understand it better.
Finally, the ultimate goal is to provide a ''replay function'', which would show how attackers proceeded and what could have been done better to prevent or mitigate attacks. 

\vspace{-0.3em}
\begin{acks}
This research was supported by the Security Research Programme of the Czech Republic 2015-2020 (BV III/1--VS) granted by the \grantsponsor{cz-interior}{Ministry of the Interior of the Czech Republic}{http://www.mvcr.cz/bezpecnostni-vyzkum.aspx} under No.~\grantnum{cz-interior}{VI20162019014} -- Simulation, detection, and mitigation of cyber threats endangering critical infrastructure.
\end{acks}

\vspace{-0.2em}
\bibliographystyle{ACM-Reference-Format}
\bibliography{sigcse2018-feedback} 

\end{document}